\documentclass[10pt]{iopart}

\begin{document}

\title{A geometric description of the intermediate behaviour for spatially
homogeneous models}
\author{Pantelis S. Apostolopoulos}

\begin{abstract}
A new approach is suggested for the study of geometric symmetries in general
relativity, leading to an invariant characterization of the evolutionary
behaviour for a class of Spatially Homogeneous (SH) vacuum and
orthogonal $\gamma -$law perfect fluid models. Exploiting the 1+3 orthonormal
frame formalism, we express the kinematical quantities of a generic symmetry
using expansion-normalized variables. In this way, a specific symmetry
assumption lead to geometric constraints that are combined with the
associated integrability conditions, coming from the existence of the
symmetry and the induced expansion-normalized form of the Einstein's Field
Equations (EFE), to give a close set of compatibility equations. By
specializing to the case of a \emph{Kinematic Conformal Symmetry} (KCS),
which is regarded as the direct generalization of the concept of
self-similarity, we give the complete set of consistency equations for the
whole SH dynamical state space. An interesting aspect of the analysis of the
consistency equations is that, \emph{at least} for class A models which are Locally Rotationally Symmetric or 
lying within the invariant subset satisfying $N_{\hspace{0.2cm}\alpha }^{\alpha
}=0 $, a proper KCS \emph{always exists} and reduces to a self-similarity of
the first or second kind at the asymptotic regimes, providing a way for the
``geometrization'' of the intermediate epoch of SH models.
\end{abstract}

\address{University of Athens, Department of Physics, Nuclear and 
Particle Physics Section, Panepistimiopolis, Zografos 15771, Athens, Greece\\ and\\
Departament de Fisica, Universitat de les Illes Balears, Cra. 
Valldemossa km 7.5, E-07122 Palma de Mallorca, Spain.}

\ead{papost@phys.uoa.gr\newline
} \relax

\maketitle

\section{Introduction}

\setcounter{equation}{0}

Due to their appealing geometric, kinematic and dynamical structure,
Spatially Homogeneous (SH or Bianchi) models have received considerable
attention in the last 3 decades. Apart from the obvious gain of a direct
generalization of the standard Friedmann-Lema\^itre cosmological model, one
of the main reasons for this interest is the fact that the Einstein's Field
Equations (EFE) are reduced to a coupled system of ordinary differential
equations. Then by introducing the orthonormal frame formalism and
expansion-normalized variables, in order to scale away the influence of the
expansion from the overall evolution of the corresponding models, one
exploits methods from the theory of dynamical systems to examine their
behaviour at early, late and intermediate periods of their history. This
approach has shown the significant role of the transitively self-similar SH
models since they represent the past and future (equilibrium) states for the
majority of evolving vacuum and $\gamma-$law perfect fluid models \cite
{Wainwright-Ellis, Coley-Book}.

However, one can supplement this discussion with a broader consideration of
the issues concerning the evolutionary era of SH models. The equilibrium
state, whenever it exists, of evolving SH models is geometrically described by a model that
admits a Homothetic Vector Field (HVF or self-similarity of the first kind)
or a Kinematic Self-Similarity (KSS or self-similarity of the second kind) $%
\mathbf{X}$ which is defined according to: 
\begin{equation}
\mathcal{L}_{\mathbf{X}}u_a=\delta u_a \qquad \mathcal{L}_{\mathbf{X}%
}h_{ab}=2\alpha h_{ab}  \label{sx1.1}
\end{equation}
where $h_{ab}=g_{ab}+u_au_b$ is the projection operator normal to the
timelike unit vector field $u^au_a=-1$ and $\alpha,\delta =$const.
essentially represent the (constant) time amplification and space dilation 
\cite{Carter Henriksen 1,Carter Henriksen 2,Coley-KSS,Carr-Coley1,Carr-Coley2}.

Therefore it will be enlightening to complete the geometric picture and find
a relevant way to invariantly (although not uniquely) characterize the
intermediate behaviour of the associated models, or to put it equivalently,
one is tempted to ask: what was the nature of the generator of the
self-similarity (either of the first or second kind) at the time of the
intermediate evolution i.e. what kind of symmetry invariantly describes the
intermediate behaviour of evolving vacuum or perfect fluid SH models? The
natural and intuitive answer (although using heuristic arguments) is that
one should expect that this type of symmetry must involve a generalization
of the self-similarity. Consequently, the symmetry will represent the smooth
transition mechanism between the evolving and the equilibrium states of SH
models i.e. its ``asymptotic behaviour'' (into the past and the future) will
be one of the well known symmetries.

In order to achieve the above goal one could consider and study general
symmetries taking into account the associated local diffeomorphisms and the
intrinsic geometric structure of the spacetime manifold \cite{Hall-Book}.
However, from a dynamical point of view, it appears natural to pursue a
different direction by making use of the full set of EFE in order to augment
and enrich the study of geometric symmetries. Clearly, this approach will
provide us the necessary set of compatibility equations which can be
studied in each subclass of SH (or even less symmetric) models. Motivated
by the above discussion and the great success of studying SH models using
elements from the theory of dynamical systems, in the present article we
propose an alternative technique to study geometric symmetries that meets
the aforementioned scope. This method fully incorporates the
expansion-normalized orthonormal frame approach and give a transparent
picture of how a specific symmetry ``assumption" can be consistently endowed
within a class of models or will produce further constraints, thus losing certain 
features of the general case. An illustrative and interesting 
example to which the preceding discussion applies is a recently presented
new type of symmetry namely the concept of \emph{Kinematic Conformal
Symmetry} (KCS). This type of symmetry is regarded as a consistent, with
geometry, generalization of the KSS or the case of conformal transformations
preserving at the same time the causal structure of the spacetime manifold 
\cite{Parrado-Senovilla1, Parrado-Senovilla2}. Therefore, as a first step
towards an invariant and geometric description of SH models, it seems
adequate to study the implications from the existence of a KCS.

The structure of the paper is as follows: in section 2, after a brief review
of the basic elements of the 1+3 orthonormal frame scheme and the
introduction of expansion-normalized variables, we specialize our study to
SH geometries and give the tetrad/expansion-normalized form of the
kinematical quantities that define a generic symmetry. The determination of
the corresponding expressions for the case of a KCS is treated in section 3.
Since any type of symmetry assumption lead to geometric constraints which
pass, through the EFE, into dynamics, we also determine the dynamical
restrictions as follow from the associated integrability conditions. This
set of equations, together with the expansion-normalized form of the EFE,
constitute a close system of compatibility equations which can be used to
check the consistent existence of a KCS in all the SH cosmologies. In
section 4, special attention is given to some models of class A in 
which we study the consistency of the system of equations. We find that, \emph{at least} models 
which are Locally Rotationally Symmetric (LRS) or lying within the invariant subset satisfying $%
N_{\hspace{0.2cm}\alpha }^{\alpha }=0$, always admit a KCS. A side result is
that, at the asymptotic regimes, the KCS reduces to HVF or a KSS which shows
that, in principle, \emph{the existence of a four dimensional group of KCS
invariantly characterizes the intermediate behaviour of the associated evolving vacuum
and perfect fluid models}. Finally in section 5 we give some concluding
remarks and comments and we discuss possible extensions and further advances
which can be made in the direction of a satisfactory and conclusive answer
to the ``geometrisation" of the evolutionary behaviour of SH (or less
symmetric) models.

Throughout this paper, the following conventions have been used: spatial frame indices
are denoted by lower Greek letters $\alpha ,\beta ,...=1,2,3$, lower Latin
letters denote spacetime indices $a,b,...=0,1,2,3$ and we use geometrized
units such that $8\pi G=c=1$.

\section{The generic symmetry in SH models}

\setcounter{equation}{0}

\subsection{The 1+3 orthonormal frame formalism}

Spatially Homogeneous models are specified in geometric terms by requiring
the existence of a $\mathcal{G}_3$ Lie algebra of Killing Vector Fields
(KVFs) $\mathbf{X}_\alpha $ with three-dimensional spacelike orbits $\mathcal{S}$%
. This implies the existence of a uniquely defined unit timelike vector
field $u^a$ ($u^au_a=-1$) normal to the spatial foliations $\mathcal{S}$:

\begin{equation}
u_{[a;b]}=0=u_{a;b}u^b\Leftrightarrow \frac 12\mathcal{L}_{\mathbf{u}%
}g_{ab}= u_{a;b}=\sigma _{ab}+\frac \theta 3h_{ab}  \label{prel1}
\end{equation}
where $\sigma _{ab},\theta =u_{a;b}g^{ab},h_{ab}=g_{ab}+u_au_b$ are the
kinematical quantities associated with the $u^a$ according to the standard
1+3 decomposition of an arbitrary timelike congruence \cite{Ellis-Elst1}.
Because $u^a$ is irrotational and geodesic, there exists a time function $%
t(x^a)$ such that $u^a=\delta _t^a$ i.e. each value of $t$ essentially
represents the hypersurfaces $\mathcal{S}$.

As far as the dynamical structure is concerned, from a cosmological point of
view, it is sufficient to focus our study on SH models with an orthogonal
perfect fluid assuming a $\gamma -$law equation of state. Therefore the EFE
become:

\begin{equation}
R_{ab}=\frac{3\gamma -2}{2}\rho u_{a}u_{b}+\frac{2-\gamma }{2}\rho h_{ab}
\label{FieldEquations}
\end{equation}
It follows that the timelike vector field $u^{a}$ is identified with the
average fluid flow velocity.

On using the orthonormal frame formalism in SH cosmologies the starting
point is to introduce a set of four linearly independent vector fields $\{%
\mathbf{e}_{a}\}$ and their dual 1-forms $\{\mathbf{\omega }^{a}\}$ which
are invariant under the three-parameter group of isometries: 
\begin{equation}
\lbrack \mathbf{X}_{\alpha },\mathbf{e}_{a}]=0,\qquad \lbrack \mathbf{X}%
_{\alpha },\mathbf{\omega }^{a}]=0.  \label{invariant1}
\end{equation}
Then, by performing a time-dependent rescaling of $\{\mathbf{e}_{a}\}$, we
can write (locally) the metric tensor in a manifestly Minkowskian form:\ 
\begin{equation}
ds^{2}=\eta _{ab}\mathbf{\omega }^{a}(t,x^\alpha)\mathbf{\omega }^{b}(t,x^\alpha).
\label{metric}
\end{equation}
In this case the invariant vector fields $\{\mathbf{e}_{a}\}$ and the
connection forms $\Gamma _{\hspace{0.2cm}bc}^{a}\mathbf{\omega }^{c}$ (where 
$\nabla _{\mathbf{e}_{c}}\mathbf{e}_{b}=\Gamma _{\hspace{0.2cm}bc}^{a}%
\mathbf{e}_{a}$) satisfy the commutation and the first Cartan structure
equations:

\begin{equation}
\lbrack \mathbf{e}_{a},\mathbf{e}_{b}]=\gamma _{\hspace{0.2cm}ab}^{c}(t)%
\mathbf{e}_{c},\qquad d\mathbf{\omega }^{a}=\Gamma _{\hspace{0.2cm}bc}^{a}%
\mathbf{\omega }^{b}\wedge \mathbf{\omega }^{c}=-\frac{1}{2}\gamma _{\hspace{%
0.2cm}bc}^{a}\mathbf{\omega }^{b}\wedge \mathbf{\omega }^{c}
\label{commutators}
\end{equation}
where $d$ and $\wedge $ are the usual exterior derivative and exterior
product of 1-forms respectively.

It follows that the commutation functions $\gamma _{ab}^{c}$ and the
connection coefficients $\Gamma _{ab}^{c}$ are related via: 
\begin{equation}
\Gamma _{abc}=\frac{1}{2}\left[ \eta _{ad}\gamma _{\hspace{0.2cm}%
cb}^{d}+\eta _{bd}\gamma _{\hspace{0.2cm}ac}^{d}-\eta _{cd}\gamma _{\hspace{%
0.2cm}ba}^{d}\right] \Leftrightarrow \gamma _{\hspace{0.2cm}bc}^{a}=-\left(
\Gamma _{\hspace{0.2cm}bc}^{a}-\Gamma _{\hspace{0.2cm}cb}^{a}\right) .
\label{connection1}
\end{equation}
Furthermore, the requirement of the constancy of the metric under covariant
differentiation implies that, $\Gamma _{(ab)c}=0$ where $\Gamma _{abc}\equiv
\eta _{da}\Gamma _{\hspace{0.2cm}bc}^{d}$.

The above definitions suggest that the tetrad form of the covariant
derivative of every tensor field is written in the well-known way \cite
{Ellis-Elst1}:

\begin{equation}
\nabla _{c}K_{\hspace{0.2cm}i...}^{a...}=\mathbf{e}_{c}\left( K_{\hspace{%
0.2cm}i...}^{a...}\right) +\Gamma _{\hspace{0.2cm}dc}^{a}K_{\hspace{0.2cm}%
i...}^{d...}+...-\Gamma _{\hspace{0.2cm}ic}^{d}K_{\hspace{0.2cm}%
d...}^{a...}-...  \label{covariantderiv}
\end{equation}
where $\mathbf{e}_{c}\left( K_{\hspace{0.2cm}i...}^{a...}\right) $ is
regarded as the directional derivative of the functions $K_{\hspace{0.2cm}%
i...}^{a...}$ along the vector field $\mathbf{e}_{c}$.

Because in SH (vacuum or perfect fluid) models there always exists the
preferred and well defined timelike vector field $u^{a}$ associated with a
congruence of curves normal to the three-dimensional spacelike orbits $\mathcal{S%
}$ of homogeneity, it is natural to select it as the timelike frame vector
i.e. $\mathbf{e}_{0}=\mathbf{u}$. It follows from equation (\ref{commutators}%
) that the kinematical quantities of the timelike congruence $\mathbf{u}$
are directly related with the commutation functions $\gamma _{\hspace{0.2cm}%
ab}^{c}$ according to \cite{Ellis-Elst1,Elst-Uggla1}:

\begin{equation}
\gamma _{\hspace{0.2cm}0\alpha }^{0}=\dot{u}_{\alpha }=0,\qquad \gamma _{%
\hspace{0.2cm}\alpha \beta }^{0}=-2\varepsilon _{\alpha \beta \gamma }\omega
^{\gamma }=0  \label{acc-vortic}
\end{equation}
\begin{equation}
\gamma _{\hspace{0.2cm}0\alpha }^{\beta }=-\frac{1}{3}\theta \delta _{%
\hspace{0.2cm}\alpha }^{\beta }-\sigma _{\hspace{0.2cm}\alpha }^{\beta
}+\varepsilon _{\hspace{0.2cm}\alpha \gamma }^{\beta }\Omega ^{\gamma }
\label{expansion-shear}
\end{equation}
where $\Omega ^{\gamma }$ is the local angular velocity of the spatial frame
with respect to a Fermi-propagated frame along $\mathbf{e}_{0}$.
On the other hand, the spatial components of $\gamma _{\hspace{0.2cm}ab}^{c}$
are decomposed as: 
\begin{equation}
\gamma _{\hspace{0.2cm}\beta \gamma }^{\alpha }=a_{\beta }\delta _{\hspace{%
0.2cm}\gamma }^{\alpha }-a_{\gamma }\delta _{\hspace{0.2cm}\beta }^{\alpha
}+\varepsilon _{\beta \gamma \rho }n^{\alpha \rho }
\label{spatialcommutator}
\end{equation}
leading to Bianchi class A and B models according to whether the quantity $%
a_{\beta }$ vanishes or not.

Combining equations (\ref{connection1}) and (\ref{acc-vortic})-(\ref
{spatialcommutator}) we easily find: 
\begin{equation}
\Gamma _{\beta 0\alpha }=\frac{\theta }{3}\delta _{\alpha \beta }+\sigma
_{\alpha \beta },\qquad \Gamma _{0\alpha 0}=0,\qquad \Gamma _{\alpha \beta
0}=\varepsilon _{\alpha \beta \gamma }\Omega ^{\gamma }  \label{connection2}
\end{equation}
\begin{equation}
\Gamma _{\alpha \beta \gamma }=2a_{[\alpha }\delta _{\beta ]\gamma
}+\varepsilon _{\gamma \rho \lbrack \alpha }n^\rho {}_{\beta ]}+\frac{1}{2}\varepsilon _{\alpha \beta \rho }n^\rho {}_{\gamma}
\label{connection3}
\end{equation}
The complete set of the gravitational field equations is expressed in terms
of the shear components $\sigma _{\alpha \beta }$, the expansion $\theta $
and the spatial curvature quantities $a_{\beta }$, $n_{\alpha \beta }$, by
utilizing the Ricci identity for $u^{a}$, the Jacobi identities for the
frame vector fields and the Bianchi identities.

\subsection{Expansion normalized variables}

Of particular importance in the exploration of the asymptotic dynamics of SH
models, is the reformulation of the complete set of orthonormal frame
equations, as an autonomous system of first order ordinary differential
equations. This can be done by defining a set of expansion-normalized
(dimensionless) variables which results the decoupling of the evolution
equation of $H=\theta /3$ from the rest of the evolution equations: 
\begin{equation}
\frac{dt}{d\tau }=\frac{1}{H},\quad \frac{dH}{d\tau }=-\left( 1+q\right) H
\label{diffequat2}
\end{equation}
where $q,H$ are the deceleration and Hubble parameter respectively and $\tau 
$ is the dimensionless time variable.

The complete set of equations can be written in the form\ \cite
{Hewitt-Bridson-Wainwright}:

\begin{equation}
\Sigma _{\alpha \beta }^{\prime }=-\left( 2-q\right) \Sigma _{\alpha \beta
}+2\epsilon _{\hspace{0.3cm}(\alpha }^{\mu \nu }\Sigma _{\beta )\mu }R_{\nu
}-S_{\alpha \beta }  \label{evol1}
\end{equation}
\begin{equation}
N_{\alpha \beta }^{\prime }=qN_{\alpha \beta }+2\Sigma _{\hspace{0.2cm}%
(\alpha }^{\mu }N_{\beta )\mu }+2\epsilon _{\hspace{0.3cm}(\alpha }^{\mu \nu
}N_{\beta )\mu }R_{\nu }  \label{evol2}
\end{equation}
\begin{equation}
A_{\alpha }^{\prime }=qA_{\alpha }-\Sigma _{\hspace{0.2cm}\alpha }^{\mu
}A_{\mu }+\epsilon _{\hspace{0.3cm}\alpha }^{\mu \nu }A_{\mu }R_{\nu }
\label{evol3}
\end{equation}
\begin{equation}
\Omega ^{\prime }=\Omega \left[ 2q-\left( 3\gamma -2\right) \right]
\label{Bianchi1}
\end{equation}
where: 
\begin{eqnarray}
S_{\alpha \beta } &=&2N_{\hspace{0.15cm}\alpha }^{\gamma }N_{\beta \gamma
}-N_{\hspace{0.15cm}\gamma }^{\gamma }N_{\alpha \beta }-\frac{1}{3}\left(
2N_{\hspace{0.15cm}\alpha }^{\gamma }N_{\beta \gamma }-N_{\hspace{0.15cm}%
\gamma }^{\gamma }N_{\alpha \beta }\right) \delta ^{\alpha \beta } 
\nonumber \\
&&-2\epsilon _{\hspace{0.3cm}(\alpha }^{\mu \nu }N_{\beta )\mu }A_{\nu }
\label{spatial-curvature}
\end{eqnarray}
and a prime denotes derivative w.r.t. $\tau $.

The above system is subjected to the algebraic constraints: 
\begin{equation}
\Omega =1-\frac{1}{6}\Sigma ^{\alpha\beta}\Sigma _{\alpha\beta}-K
\label{algebraic1}
\end{equation}
\begin{equation}
3\Sigma _{\hspace{0.2cm}\alpha }^{\beta }A_{\beta }-\epsilon _{\alpha }^{%
\hspace{0.2cm}\mu \nu }\Sigma _{\hspace{0.2cm}\mu }^{\beta }N_{\beta \nu }=0
\label{algebraic2}
\end{equation}
where: 
\begin{equation}
K=\frac{1}{12}\left( 2N_{\hspace{0.15cm}\alpha }^{\gamma }N_{\beta \gamma
}-N_{\hspace{0.15cm}\gamma }^{\gamma }N_{\alpha \beta }\right) \delta
^{\alpha \beta }+A_{\gamma }A^{\gamma }  \label{curvature1}
\end{equation}
and the deceleration parameter is given by the relation: 
\begin{eqnarray}
q &=&\frac{1}{3}\Sigma ^{\alpha\beta}\Sigma _{\alpha\beta}+\frac{1}{2}\Omega %
\left[ \left( 3\gamma -2\right) \right]   \nonumber \\
&=&2\left( 1-K\right) +\frac{1}{2}\Omega \left[ 3\left( \gamma -2\right) %
\right] .  \label{algebraic4}
\end{eqnarray}

\subsection{The generic symmetry in expansion-normalized variables}

The folklore for investigating the implications of the existence of
geometric symmetries in General Relativity can be divided into two main
categories. The first category is a geometric approach in which we study symmetries taking into
account the holonomy group structure of the spacetime manifold together with
the associated local diffeomorphisms \cite{Hall-Book}. In the second category one 
formulates the necessary and sufficient conditions, coming from the existence
of the symmetry, in a covariant way and study their consequences in the
kinematics and dynamics of the corresponding model \cite{Tsamparlis1}. Of
course one could also deal directly with the resulting system of partial
differential equations (pdes), presupposing a specific geometrical and
dynamical configuration which render the symmetry equations to be more
tractable. Obviously this approach undergo many disadvantages and
pathologies. One of the serious stumbling blocks is the fact that as the
generality of a model is increased (i.e. the underlying geometric structure of the model 
is less symmetric than the SH geometry) the symmetry pdes are progressively
non-linear and very often lead to solutions of the EFE which are immediately ruled out physically.

Here we suggest an alternative approach for the study of geometric
symmetries which fully exploits the well-established orthonormal frame
formalism in terms of expansion-normalized variables. Although this
technique will be applied to a specific symmetry ``assumption" (as we shall
see in the next section this is not really an assumption, at least for a 
class of models, but a consequence of the complete set of EFE) it can 
be used in a more general context for the study of other types of important
symmetries \cite{Apostol-Carot1}.

Let us consider an arbitrary vector field $\mathbf{X}$ and express its
components in terms of the frame vector fields $\mathbf{e}_a$: 
\begin{equation}
\mathbf{X}=X^{a}\mathbf{e}_{a}=\lambda \mathbf{e}_{0}+X^{\alpha }\mathbf{e}%
_{\alpha }  \label{symmetryvector}
\end{equation}
where $\lambda ,X^{\alpha }$ are continuously differential functions of the
spacetime manifold.

The first derivatives of $\mathbf{X}$ are decomposed into irreducible
symmetry kinematical parts $\{\psi ,H_{ab},F_{ab}\}$ in the standard way: 
\begin{equation}
\nabla _{b}X_{a}=\psi g _{ab}+H_{ab}+F_{ba}  \label{kinematicalparts}
\end{equation}
where 
\begin{equation}
4\psi \equiv \nabla _{k}X^{k},  \label{conformalfactor1}
\end{equation}
\begin{equation}
H_{ab}=\left[ \nabla _{(b}X_{a)}-\frac{1}{4}\left( \nabla _{k}X^{k}\right) g
_{ab}\right]  \label{tracelesspart}
\end{equation}
\begin{equation}
F_{ab}=-\nabla _{\lbrack b}X_{a]}  \label{antisymmetric-part}
\end{equation}
are the conformal factor, the traceless symmetric part and the antisymmetric
part respectively.

Using the definition (\ref{covariantderiv}) and equations (\ref{connection2}%
), (\ref{connection3}), (\ref{symmetryvector})-(\ref{antisymmetric-part}),
we find the tetrad analogue of the kinematical quantities. If we further
invoke the expansion-normalized differential operators $\mathbf{\partial }%
_{a}\equiv \mathbf{e}_{a}/H$ in the general expressions we finally obtain:

\begin{equation}
4\psi =H\left[ \mathbf{\partial }_{0}\left( \lambda \right) +\mathbf{%
\partial }_{\alpha }(X^{\alpha })+3\lambda -2A_{\alpha }X^{\alpha }\right]
\label{conformalfactor}
\end{equation}

\begin{equation}
H_{00}=\frac{H}{4}\left[ -3\mathbf{\partial }_{0}(\lambda )+\mathbf{\partial 
}_{\alpha }(X^{\alpha })+3\lambda -2A_{\alpha }X^{\alpha }\right]
\label{H00}
\end{equation}
\begin{equation}
2H_{0\alpha }=H\left[ -\mathbf{\partial }_{\alpha }(\lambda )+\mathbf{%
\partial }_{0}(X_{\alpha })-\left( \delta _{\alpha \gamma }+\Sigma _{\alpha
\gamma }\right) X^{\gamma }+\varepsilon _{\alpha \beta \gamma }R^{\gamma
}X^{\beta }\right]  \label{H0-alpha}
\end{equation}
\begin{eqnarray}
H_{\alpha \beta } &=&H\{\mathbf{\partial }_{(\beta }X_{\alpha )}+\left(
\delta _{\alpha \beta }+\Sigma _{\alpha \beta }\right) \lambda -\left[
A_{\gamma }\delta _{\alpha \beta }-A_{(\alpha }\delta _{\beta )\gamma }%
\right] X^{\gamma } \nonumber \\
&&+\varepsilon _{\gamma \delta (\alpha }N_{\hspace{0.2cm}\beta )}^{\delta
}X^{\gamma }\}-\psi \delta _{\alpha \beta }
\label{Halpha-beta}
\end{eqnarray}
\begin{equation}
2F_{0\alpha }=H\left[ \mathbf{\partial }_{\alpha }(\lambda )+\mathbf{%
\partial }_{0}(X_{\alpha })+\left( \delta _{\alpha \gamma }+\Sigma _{\alpha
\gamma }\right) X^{\gamma }-\varepsilon _{\alpha \beta \gamma }R^{\gamma
}X^{\beta }\right]  \label{F0-alpha}
\end{equation}
\begin{equation}
F_{\alpha \beta }=H\left\{ -\mathbf{\partial }_{[\beta }X_{\alpha ]}-\left[
A_{[\alpha }\delta _{\beta ]\gamma }+\frac{1}{2}\varepsilon _{\alpha \beta
\delta }N_{\hspace{0.2cm}\gamma }^{\delta }\right] X^{\gamma }\right\} .
\label{Falpha-beta}
\end{equation}
Equations (\ref{conformalfactor})-(\ref{Falpha-beta}) represent the \emph{%
symmetry kinematical quantities in terms of expansion-normalized variables}
in SH geometries and can be used in order to have a first hint of how the
dynamics affects on the geometry of SH models (or vice-versa). We note that,
one could choose to define expansion-normalized symmetry kinematical
quantities in a similar way we do for the shear and spatial curvature
variables. This would be convenient for high symmetries where first or
second derivatives of $\{\psi ,H_{ab},F_{ab}\}$ are involved. However for
the cases we are interested in the present article the use of (\ref
{conformalfactor})-(\ref{Falpha-beta}) will be satisfactory.

We conclude this section by pointing out that, because any type of symmetry
assumption is described in terms of geometric constraints, these are
inherited by the dynamics through the EFE (\ref{FieldEquations}). Therefore
in order to visualize how the symmetry further interacts with the dynamics,
it is necessary to determine the effect of the former on the Ricci tensor.
By using the well-known commutation relation between the connection and the
Lie derivative \cite{Yano}: 
\[
\mathcal{L}_{\mathbf{X}}\Gamma _{\hspace{0.2cm}bc}^{a}=\frac{1}{2}g^{ar}%
\left[ \nabla _{c}\left( \mathcal{L}_{\mathbf{X}}g_{br}\right) +\nabla
_{b}\left( \mathcal{L}_{\mathbf{X}}g_{cr}\right) -\nabla _{r}\left( \mathcal{%
L}_{\mathbf{X}}g_{bc}\right) \right] 
\]
\[
\mathcal{L}_{\mathbf{X}}R_{ab}=\nabla _{c}\left[ \Gamma _{\hspace{0.2cm}%
\left( ab\right) }^{c}\right] -\nabla _{(b}\left[ \Gamma^c {} _{a)c}\right] 
\]
and the defining equation (\ref{kinematicalparts}), we can show after a
straightforward calculation that:\ 
\begin{equation}
\mathcal{L}_{\mathbf{X}}R_{ab}=-2\nabla _{b}\nabla _{a}\psi -g_{ab}\nabla
_{c}\nabla ^{c}\psi +2\nabla _{k}\nabla _{(b}H^k {}_{a)}-\nabla
_{c}\nabla ^{c}H_{ab}.  \label{LieRicci1}
\end{equation}
Essentially, the last equation represents a set of integrability conditions
which can be used to check the consistent existence of \emph{every type} of
symmetry assumption.

\section{Generalized Conformal Symmetries in SH models}

\setcounter{equation}{0} Although there exists (and can be defined) a
sufficiently large number of symmetries, the most important type of them (up
to date) appears to concern the constant scale invariance of the metric
represented by the existence of a proper HVF \cite{Carr-Coley1,Carr-Coley2,Carot-Sintes}.
For SH vacuum and perfect fluid models this is indeed the case due to the
profound relevance of homothetic models with the equilibrium points of the
SH state space \cite{Hsu-Wainwright,Apostol9,Apostol15,Apostol16}.

Recently, a new type of symmetry has been suggested, the so-called \emph{%
bi-conformal transformations} which can be interpreted as generalizing the
concepts of the self-similarity and the conformal motions. In the present
work we will concern with an interesting subcase, that is, the so-called 
\emph{Kinematic Conformal Symmetry} (KCS). In particular a smooth vector
field $\mathbf{X}$ is the generator of a KCS \emph{iff} the following
relations hold \cite{Parrado-Senovilla1}: 
\begin{equation}
\mathcal{L}_{\mathbf{X}}u_{a}=\delta u_{a},\qquad \mathcal{L}_{\mathbf{X}%
}h_{ab}=2\alpha h_{ab}  \label{definitionKCS}
\end{equation}
or, in terms of the metric: 
\begin{equation}
\mathcal{L}_{\mathbf{X}}g_{ab}=2\alpha g_{ab}+2\left( \alpha -\delta \right)
u_{a}u_{b}  \label{KCSwithmetric}
\end{equation}
where $\alpha ,\delta $ are smooth functions that we shall call \emph{%
symmetry scales} and $h_{ab}=g_{ab}+u_{a}u_{b}$ is the projection operator
perpendicular to the timelike congruence $u_{a}$.

Combining equations (\ref{kinematicalparts}) and (\ref{KCSwithmetric}) we
express the symmetry kinematical parts in the form: 
\begin{equation}
\psi =\frac{3\alpha +\delta }{4},\qquad H_{ab}=\frac{\alpha -\delta }{4}%
\left( g_{ab}+4u_{a}u_{b}\right) .  \label{KCSkinematical}
\end{equation}
It can be easily observed that when $\alpha =\delta $ the KCS reduces to a
Conformal Vector Field (CVF) which, due to the equation (\ref{definitionKCS}%
), is necessarily inheriting i.e. the integral curves of $u^{a}$ are mapped
conformally by the CVF $\mathbf{X}$ \cite{Coley-Hall-Keane-Tupper}. As a
result the Lie algebra $\mathcal{I} $ of inheriting CVFs is always a subalgebra of the
Lie algebra of KCS which shall be denoted as $\mathcal{B}$. Moreover when
the symmetry scales $\alpha ,\delta $ are both (different) constants we
recover the case of a Kinematic Self-Similarity.

\subsection{Symmetry constraints}

Clearly there exists a direct dependence between the existence of a KCS, as
well as any other type of symmetry assumption, and a specific cosmological
model. This mutual influence is reflected in the induced geometric,
kinematic and dynamic constraints which are imposed due to the intrinsic
nature of the symmetry and/or as a consequence of the specific geometric and
dynamical structure of the spacetime manifold. In the case of a KCS these
constraints are derived from the general relations (\ref{conformalfactor})-(%
\ref{Halpha-beta}) and the symmetry assumptions (\ref{KCSkinematical}): 
\begin{equation}
3\alpha= H\left[\mathbf{\partial }_{\alpha }(X^{\alpha })+3\lambda -2A_{\alpha }X^{\alpha }%
\right]  \label{conformconstraint}
\end{equation}

\begin{equation}
\left[ -4H\mathbf{\partial }_{0}(\lambda )+3\alpha +\delta \right] =3\left(
\alpha -\delta \right)\Rightarrow \delta=H\mathbf{\partial }_{0}(\lambda )  \label{H00constraint}
\end{equation}
\begin{equation}
-\mathbf{\partial }_{\alpha }(\lambda )+\mathbf{\partial }_{0}(X_{\alpha
})-\left( \delta _{\alpha \gamma }+\Sigma _{\alpha \gamma }\right) X^{\gamma
}+\varepsilon _{\alpha \beta \gamma }R^{\gamma }X^{\beta }=0
\label{H0aconstraint}
\end{equation}
\begin{eqnarray}
&&H\{\mathbf{\partial }_{(\beta }X_{\alpha )}+\left( \delta _{\alpha \beta
}+\Sigma _{\alpha \beta }\right) \lambda -\left[ A_{\gamma }\delta _{\alpha
\beta }-A_{(\alpha }\delta _{\beta )\gamma }\right] X^{\gamma }  \nonumber
\\
&&+\varepsilon _{\gamma \delta (\alpha }N_{\hspace{0.2cm}\beta )}^{\delta
}X^{\gamma }\}-\alpha \delta _{\alpha \beta }=0.  \label{Habconstraint}
\end{eqnarray}
We should emphasize that the above set of constraints must be augmented with
the associated Jacobi identities satisfied by the generators $\left\{ 
\mathbf{X}_{\alpha },\mathbf{X}\right\} $ that constitute the Lie Algebra $%
\mathcal{B}$ of KCS. This will require the determination of the dimension of 
$\mathcal{B}$ and the assumption that the former is finite since there are
cases for which $\mathcal{B}$ is infinite dimensional \cite
{Parrado-Senovilla1}. Nevertheless we shall not pursue the problem in full
generality and we will restrict our considerations to the case where the
dimension of $\mathcal{B}$ is finite and equal to four (together with the $%
\mathcal{G}_{3}$ of KVFs which can be seen as ``trivial'' KCS). In fact this
assumption seems reasonable and not even restrictive, from a geometrical
point of view, because we intend to describe the ``intermediate behaviour''
of a (proper) self-similarity of the first or second kind in a way that is
identified with the presence of a KCS. In this case it can be shown from
Jacobi identities \cite{Petrov} that the Lie bracket of a KCS with the KVFs
is always a linear combination of the later. It turns out that the scalar $%
\lambda $ and the symmetry scales $\alpha ,\delta $ (due to equation (\ref{definitionKCS})) 
are spatially homogeneous: 
\begin{equation}
\partial _{\alpha }\lambda =\partial _{\alpha }\alpha =\partial _{\alpha
}\delta =0.  \label{spatialhomofscales}
\end{equation}
In addition we point out that equation (\ref{KCSwithmetric}) is a
consequence of (\ref{definitionKCS}) since the former suppresses the scale
amplification of both time and space which is essentially represented by the
latter. However the spatial homogeneity condition (\ref{spatialhomofscales}%
) ensures that the definitions (\ref{KCSwithmetric}) and (\ref{definitionKCS}%
) are equivalent.

\subsection{Integrability conditions}

Due to the geometric character of the KCS to preserve the causal structure
of the spacetime manifold, it is natural to expect that the associated
transformation group will respect, to some level, the intrinsic properties 
of SH models. Therefore one should expect that the presence of a
KCS will induce weaker constraints than the case of CVFs or the KSS. In
order to confirm this, it will be required to determine the integrability
conditions adapted to the case of SH models filled, in general, with a $%
\gamma -$law non-tilted perfect fluid.

As we have shown, the scale functions and the trace $\psi $ are
both spatially homogeneous which implies that equation (\ref{LieRicci1}),
due to (\ref{KCSkinematical}), can be written:\newline
\begin{eqnarray}
\mathcal{L}_{\mathbf{X}}R_{ab} &=&3\left[ H\left( \dot{\delta}-2\dot{\alpha}%
\right) -\ddot{\alpha}\right] u_{a}u_{b}  \nonumber \\
&&+\left[ 2\left( \alpha -\delta \right) \left( 3H^{2}+\dot{H}\right) +\ddot{%
\alpha}+H\left( 6\dot{\alpha}-\dot{\delta}\right) \right] h_{ab}  \nonumber
\\
&&+\left[ \left( 3\dot{\alpha}-\dot{\delta}\right) \sigma _{ab}+\left(
\alpha -\delta \right) \left( 6H\sigma _{ab}+2\dot{\sigma}_{ab}\right) %
\right]  \label{LieRicci2}
\end{eqnarray}
where a dot ``$\cdot $'' denotes differentiation w.r.t. $u^{a}$.

On the other hand the EFE (\ref{FieldEquations}) and the relations (\ref
{definitionKCS}) give:\newline
\begin{equation}
\mathcal{L}_{\mathbf{X}}R_{ab}=\frac{3\gamma -2}{2}\left[ \mathbf{X}\left(
\rho \right) +2\delta \rho \right] u_{a}u_{b}+\frac{2-\gamma }{2}\left[ 
\mathbf{X}\left( \rho \right) +2\alpha \rho \right] h_{ab}.
\label{LieRicci3} \\
\end{equation}
The complete set of integrability conditions, coming from the existence of a
KCS, follow by equating (\ref{LieRicci2}) and (\ref{LieRicci3}): 
\begin{equation}
3\left[ H\left( \dot{\delta}-2\dot{\alpha}\right) -\ddot{\alpha}\right] =%
\frac{3\gamma -2}{2}\left[ \mathbf{X}\left( \rho \right) +2\delta \rho %
\right]  \label{integrability1}
\end{equation}
\begin{equation}
2\left( \alpha -\delta \right) \left( 3H^{2}+\dot{H}\right) +\ddot{\alpha}%
+H\left( 6\dot{\alpha}-\dot{\delta}\right) =\frac{2-\gamma }{2}\left[ 
\mathbf{X}\left( \rho \right) +2\alpha \rho \right]  \label{integrability2}
\end{equation}
\begin{equation}
\left( 3\dot{\alpha}-\dot{\delta}\right) \sigma _{ab}+\left( \alpha -\delta
\right) \left( 6H\sigma _{ab}+2\dot{\sigma}_{ab}\right) =0.
\label{integrability3}
\end{equation}
We observe that in the case of a CVF where $\alpha =\delta $, the last
equation implies the well-known result $\dot{\alpha}\sigma _{ab}=0$ i.e.
either the (necessary inheriting) CVF reduces to a HVF or the spacetime is
the Robertson-Walker spacetime \cite{Coley-Tupper1}.

The system of equations (\ref{integrability1})-(\ref{integrability3}) can be
conveniently reformulated in expansion-normalized variables in order to
append them in the autonomous set (\ref{evol1})-(\ref{Bianchi1}) and the
symmetry equations (\ref{conformconstraint})-(\ref{Habconstraint}). For
simplification purposes we define the dimensionless symmetry scale
functions: 
\begin{equation}
\tilde{\alpha}=\frac{\alpha }{H},\qquad \tilde{\delta}=\frac{\delta }{H}.
\end{equation}
Then, taking into account equations (\ref{evol1}) and (\ref{Bianchi1}), an
appropriate combination of (\ref{integrability1})-(\ref{integrability3})
eliminates the second order time-derivatives and give evolution equations
for $\tilde{\alpha}$ and $\tilde{\delta}$:\newline
\begin{eqnarray}
4\left[ \tilde{\alpha}^{\prime }-\left( q+1\right) \tilde{\alpha}\right]
&=&-2\left( \tilde{\alpha}-\tilde{\delta}\right) \left( 2-q\right) 
\nonumber \\
&&+\left[ \left( 3\gamma -2\right) \tilde{\delta}+3\left( 2-\gamma \right) 
\tilde{\alpha}-6\lambda \gamma \right] \Omega  \label{evolutalpha}
\end{eqnarray}
\begin{equation}
\left[ \left( 3\tilde{\alpha}-\tilde{\delta}\right) ^{\prime }-\left(
q+1\right) \left( 3\tilde{\alpha}-\tilde{\delta}\right) \right] \Sigma
_{\alpha \beta }-2\left( \tilde{\alpha}-\tilde{\delta}\right) S_{\alpha
\beta }=0.  \label{evolutdelta}
\end{equation}
\vspace{0.5cm} It is interesting to note that equation (\ref{evolutdelta}) 
\emph{excludes the existence of a proper KSS in Bianchi vacuum or
perfect fluid cosmologies with $S_{\alpha \beta}\neq 0$}. The constraint $S_{\alpha \beta}= 0$ is identically 
satisfied in Kasner type I models which is well known to admit a (proper) self-similarity of the second kind \cite
{Coley-KSS,Apostol-Tsampa4}. This result implies that a KSS fails to be
considered as a generic candidate to describe the intermediate behaviour of general
SH models, but rather one should explore the possibility for a symmetry,
representing a direct generalization of the conformal motions, in such a way
that its asymptotic behaviour is the self-similarity of the first or second
kind. We shall demonstrate in the next section that the case of a KCS
provides an evidence towards a satisfactory (but not conclusive) answer to this question, for a
significant subclass of evolving SH vacuum and perfect fluid models.

\section{Application to SH models of class A}

\setcounter{equation}{0}

Spatially homogeneous cosmologies of class A are defined by the condition $%
A^{\alpha }=0$, which due to equation (\ref{algebraic2}), implies that the
shear and spatial curvature matrices $\Sigma _{\alpha \beta },N_{\alpha
\beta }$ commute. Therefore $\Sigma _{\alpha \beta },N_{\alpha \beta }$ have
a common eigenframe and we can write: 
\begin{equation}
\Sigma _{\alpha \beta }=\mbox{diag}\left( \Sigma _{11},\Sigma _{22},\Sigma
_{22}\right) ,\qquad N_{\alpha \beta }=\mbox{diag}\left(
N_{11},N_{22},N_{22}\right) .  \label{shear-spatialcurvature}
\end{equation}
Furthermore the evolution equations (\ref{evol1}), (\ref{evol2}) and
equation (\ref{shear-spatialcurvature}) show that $R^{\alpha }=0$ which
means that the common eigenframe of $\Sigma _{\alpha \beta },N_{\alpha \beta
}$ is also Fermi-propagated. Under these conditions the study of the class A
models is considerably simplified and permits one to investigate the set of
consistency equations constituting of the EFE (\ref{evol1})-(\ref{Bianchi1}%
), the geometric constraints (\ref{conformconstraint})-(\ref{Habconstraint})
and the integrability conditions (\ref{evolutalpha})-(\ref{evolutdelta}) in
a straightforward way. To illustrate the method that could be used for the
consistency checking we outline some applications by restricting our study
to type I, II and VI$_0$ models.

However, before we proceed, it will be useful to give the corresponding
analysis for the flat Friedmann-Lema\^itre model $\mathcal{F}$ since it
represents the past or future attractor for several SH models of class A.

\subsection{Flat Friedmann-Lema\^{i}tre $\protect\gamma -$law perfect fluid
models}

It is well known that the (flat) Robertson-Walker spacetime: 
\begin{equation}
ds^{2}=-dt^{2}+S^{2}(t)\left( dx^{2}+dy^{2}+dz^{2}\right)  \label{FRW-}
\end{equation}
has a variety of ways for an invariant characterization of its structure.
For example, kinematically, is defined by the vanishing of the shear,
vorticity and acceleration of the preferred timelike congruence $u^{a}$
which, due to equations (\ref{algebraic1}) and (\ref{algebraic4}), implies:\ 

\begin{equation}
\Omega =1,\qquad q=\frac{3\gamma -2}{2}  \label{FRW1}
\end{equation}
On the other hand one can use geometric terms and describe the
Friedmann-Lema\^{i}tre model by the existence of nine proper CVFs, one of
which is parallel to $u^{a}$ \cite{Maartens-Maharaj1,Tsamp1}. Although in the case
of a $\gamma -$law perfect fluid model a proper HVF and KSS exists \cite
{Coley-KSS} (lowering the dimension of the Lie algebra of conformal motions
to eight) this does not mean that we have exhausted all the possible
(geometric) ways for the description of Friedmann-Lema\^{i}tre models. As a
result one should expect that a KCS will also exists without imposing extra
geometrical or dynamical restrictions (which is often the case for other
types of symmetries). Indeed from the constraints (\ref{conformconstraint})-(%
\ref{Habconstraint}) we find: 
\begin{equation}
3(\tilde{\alpha}-\lambda )=\mathbf{\partial }_{\alpha }(X^{\alpha }),\qquad 
\tilde{\delta}=\lambda ^{\prime }  \label{FRW2}
\end{equation}
\begin{equation}
\mathbf{\partial }_{0}(X_{\alpha })-X_{\alpha }=0,\qquad \mathbf{\partial }%
_{(\beta }X_{\alpha )}+\left( \lambda -\tilde{\alpha}\right) \delta _{\alpha
\beta }=0.  \label{FRW3}
\end{equation}
Using the set of equations (\ref{FRW1})-(\ref{FRW3}) and the integrability
condition (\ref{evolutalpha}) we can show that a KCS \emph{always exists} in
a Friedmann-Lema\^{i}tre model and the symmetry scale $\tilde{\alpha}$ is
given in terms of the temporal component:

\begin{equation}
\left( \lambda -\tilde{\alpha}\right) ^{\prime }=\left( q+1\right) \left(
\lambda -\tilde{\alpha}\right) =\frac{3\gamma }{2}\left( \lambda -\tilde{%
\alpha}\right) .  \label{FRW4}
\end{equation}
From the above equations we determine the exact form of the KCS in
the Robertson-Walker spacetime:\ 
\begin{equation}
\mathbf{X}=\lambda (t)\partial _{t}+c\left( x\partial _{x}+y\partial
_{y}+z\partial _{z}\right)  \label{FRW5}
\end{equation}
where: 
\begin{equation}
\alpha (t)=\frac{2\lambda (t)+3c\gamma t}{3t\gamma }  \label{FRW6}
\end{equation}
and the scale factor is given by $S(t)=t^{2/3\gamma }$. We note that, after a suitable change of the basis 
of the KCS Lie algebra, we can set the constant $c=0$ which implies that $\mathbf{X}$ is also 
parallel to the fluid velocity $u^a$.    

\subsection{Type I models}

In this case $N_{\alpha \beta }=0=S_{\alpha \beta }$ and equation (\ref
{evolutdelta}) gives: 
\begin{equation}
3\tilde{\alpha}-\tilde{\delta}=\tilde{c}\equiv \frac{c}{H}\Leftrightarrow
3\alpha -\delta =c  \label{typeI1}
\end{equation}
where $c$ is an arbitrary constant.

In addition, equation (\ref{H00constraint}) implies that: 
\begin{equation}
\tilde{\delta}=\lambda ^{\prime }.  \label{typeI2}
\end{equation}
We should pointed out that the symmetry constraints are necessary to ensure
the existence of a KCS. However this does not imply that they will be
preserved along the integral curves of the timelike vector field $\mathbf{e}%
_{0}$. Therefore we must propagate equations (\ref{conformconstraint})-(\ref
{Habconstraint}) in order to retain the existence of a KCS in \emph{every}
spacelike hypersurface $\mathcal{S}$. After a short calculation and the use
of the commutator relations (\ref{expansion-shear}), we obtain:\ 
\begin{equation}
\lambda ^{\prime }-\left( q+1\right) \lambda =\tilde{\alpha}^{\prime
}-\left( q+1\right) \tilde{\alpha}  \label{typeI3}
\end{equation}
\begin{equation}
\left[ \left( q-2\right) \lambda +\lambda ^{\prime }-\left( q+1\right)
\lambda \right] \Sigma _{\alpha \beta }+\Lambda _{\alpha \beta }=0
\label{typeI4}
\end{equation}
where 
\begin{equation}
\Lambda _{\alpha \beta }=\Sigma _{\gamma (\alpha }X_{\hspace{0.2cm},\beta
)}^{\gamma }-\Sigma _{\gamma (\alpha }X_{\beta )} {}^{,\gamma }
\label{typeI5}
\end{equation}
and $X_{\hspace{0.2cm},\beta }^{\gamma }\equiv \partial _{\beta }X^{\gamma }$%
.

From the $\alpha \alpha -$component of (\ref{typeI4}) it follows that: 
\begin{equation}
\left( q-2\right) \lambda +\lambda ^{\prime }-\left( q+1\right) \lambda
=0\Leftrightarrow \lambda ^{\prime }=3\lambda  \label{typeI6}
\end{equation}
while the equation $\Lambda _{\alpha \beta }=0$ expresses the spatial
components of the KCS in terms of the shear variables.

In summary we have shown that \emph{every type I }$\gamma -$\emph{law
perfect fluid model always admits a four dimensional group of Kinematic
Conformal Symmetries}. An interesting feature of this result concerns the
past ``asymptotic behaviour'' of the KCS. In particular, from equation (\ref
{typeI6}) we observe that $\lambda ^{\prime }-\left( q+1\right) \lambda
=\left( 2-q\right) \lambda $ hence at the equilibrium point ($q=2$) we have $%
\lambda ^{\prime }-\left( q+1\right) \lambda =0$ i.e. $\alpha ,\delta =$%
const. and the KCS reduces to a proper KSS with $\lambda =\mbox{const.}%
\times H^{-1}$. The future state of the KCS is treated similarly. The type I
models, approach at late times the Friedmann-Lema\^itre model $\mathcal{F}$%
, so equation (\ref{FRW4}) is trivially satisfied and the temporal component
is given in equation (\ref{typeI6}). This implies that $\alpha^{\prime
}-\left( q+1\right) \alpha \neq 0 \Rightarrow \alpha,\delta\neq$const. as
expected. Therefore, one could argue that \emph{the existence of a proper
KCS describes, in a geometric fashion, the intermediate behaviour of the
evolving type I models}.

We conclude the type I case by giving the exact form of the KCS in local
coordinates. The general $\gamma -$law perfect fluid solution can be
conveniently written in the form \cite{Wainwright-Ellis}:\ 
\begin{equation}
ds^{2}=-A^{2(\gamma -1)}dt^{2}+\sum\limits_{\alpha }t^{2p_{\alpha
}}A^{2\left( 2/3-p_{\alpha }\right) }\left( dx^{\alpha }\right) ^{2}
\label{typeIperfect1}
\end{equation}
where the function $A^{2-\gamma }=k+m^{2}t^{2-\gamma }$ and $k,m$ are
constants.

It follows that the (local) coordinate form of the KCS is:\ 
\begin{equation}
\mathbf{X}=\lambda (t)\partial _{t}+\left( ck+c_{3}\right) x\partial
_{x}+c_{2}y\partial _{y}+c_{3}z\partial _{z}  \label{KCSform}
\end{equation}
where: 
\begin{equation}
\lambda (t)=\frac{tkcA}{\left( p_{2}+2p_{3}-1\right) \left( A-tA_{,t}\right) 
}  \label{lambdaexpression}
\end{equation}
and the constant $c_{2}$ is given by: 
\begin{equation}
c_{2}=\frac{c_{3}\left( p_{2}+2p_{3}-1\right) +kc\left( p_{3}-p_{2}\right) }{%
p_{2}+2p_{3}-1}  \label{constanttypeI}
\end{equation}
where we have used the well-known relations: 
\begin{equation}
p_{1}+p_{2}+p_{3}=1,\qquad p_{1}^{2}+p_{2}^{2}+p_{3}^{2}=1
\label{KasnerExponents}
\end{equation}
satisfied by the Kasner exponents $p_{\alpha }$.

We also note that, at early times $\lambda (t)\approx t$ and $\alpha ,\delta
=$const. whereas at late times, $k=0$ and $\alpha ,\delta \neq$const. which
confirm the reduction of the KCS to a proper KSS \cite{Apostol-Tsampa4} and
to the KCS (\ref{FRW5}) respectively.

\subsection{Type II models}

The Bianchi type II invariant set is characterized by the conditions $%
N_{1}>0 $ and $N_{2}=N_{3}=0$. We find convenient to collect the consistency
equations as follow from (\ref{evol1})-(\ref{Bianchi1}), (\ref
{conformconstraint})-(\ref{Habconstraint}) and (\ref{evolutalpha})-(\ref
{evolutdelta}):

\begin{equation}
\lambda ^{\prime }-\left( q+1\right) \lambda =\tilde{\alpha}^{\prime
}-\left( q+1\right) \tilde{\alpha}  \label{TypeII-0}
\end{equation}
\begin{eqnarray}
0 &=&\left( \lambda ^{\prime }-3\lambda \right) \Sigma _{\alpha \beta
}+\Lambda _{\alpha \beta }+\left[ 2\Sigma _{11}X^{\gamma }+\Sigma _{\hspace{%
0.2cm}\varepsilon }^{\gamma }X^{\varepsilon }\right] N_{1}\varepsilon
_{\gamma 1(\alpha }\delta _{\beta )}^{1}  \nonumber \\
&&-\left( \frac{2}{3}\delta _{\alpha }^{1}\delta _{\beta }^{1}-{\frac{1}{3}}%
\delta _{\alpha }^{2}\delta _{\beta }^{2}-\frac{1}{3}\delta _{\alpha
}^{3}\delta _{\beta }^{3}\right) N_{1}^{2}\lambda  \label{TypeII-1}
\end{eqnarray}
\begin{eqnarray}
4\left[ \tilde{\alpha}^{\prime }-\left( q+1\right) \tilde{\alpha}\right]
&=&-2\left( \tilde{\alpha}-\tilde{\delta}\right) \left( 2-q\right)  
\nonumber \\
&&+\left[ \left( 3\gamma -2\right) \tilde{\delta}+3\left( 2-\gamma \right) 
\tilde{\alpha}-6\lambda \gamma \right] \Omega  \label{TypeII-11}
\end{eqnarray}
\begin{eqnarray}
0 &=&\left[ \left( 3\tilde{\alpha}-\tilde{\delta}\right) ^{\prime }-\left(
1+q\right) \left( 3\tilde{\alpha}-\tilde{\delta}\right) \right] \Sigma
_{\alpha \beta }  \nonumber \\
&&-\left( \frac{2}{3}\delta _{\alpha }^{1}\delta _{\beta }^{1}-\frac{1}{3}%
\delta _{\alpha }^{2}\delta _{\beta }^{2}-\frac{1}{3}\delta _{\alpha
}^{3}\delta _{\beta }^{3}\right) N_{1}^{2}\left( {\tilde{\alpha}-\tilde{%
\delta}}\right)  \label{TypeII-2}
\end{eqnarray}
\begin{equation}
\tilde{\delta}=\lambda ^{\prime }  \label{TypeII-2a}
\end{equation}
\begin{equation}
1-\frac{1}{{12}}N_{1}^{2}-\Sigma ^{2}-\Omega =0,\qquad q=2\Sigma ^{2}+\frac{%
3\gamma -2}{2}\Omega  \label{TypeII-3}
\end{equation}
\begin{equation}
\Sigma _{\alpha \beta }^{\prime }=(q-2)\Sigma _{\alpha \beta }-\left( \frac{2%
}{3}\delta _{\alpha }^{1}\delta _{\beta }^{1}-\frac{1}{3}\delta _{\alpha
}^{2}\delta _{\beta }^{2}-\frac{1}{3}\delta _{\alpha }^{3}\delta _{\beta
}^{3}\right) N_{1}^{2}  \label{TypeII-4}
\end{equation}
\begin{equation}
N_{1}^{\prime }=\left( q+2\Sigma _{11}\right) N_{1}.  \label{TypeII-5}
\end{equation}
Note that, in complete analogy with the type I models, equations (\ref
{TypeII-0}) and (\ref{TypeII-1}) are the result of the propagation of the symmetry constraints (\ref{conformconstraint}) and (\ref
{Habconstraint}) along $%
\partial_0$.

A quick observation can be made, due to the form of (\ref{TypeII-1}) or (\ref
{TypeII-2}). For example, the first equation implies\footnote{%
We recall the traceless property of $\Sigma _{\alpha \beta }$ i.e. $\Sigma
_{11}+\Sigma _{22}+\Sigma _{33}=0$.}: 
\begin{equation}
\left( \lambda ^{\prime }-3\lambda \right) \Sigma _{22}=\left( \lambda
^{\prime }-3\lambda \right) \Sigma _{33}=-\frac{1}{3}N_{1}^{2}\lambda
\label{TypeII6}
\end{equation}
where for $\Sigma _{22}=0$ the KCS turns to the KVF of Bianchi type II
models.

Equation (\ref{TypeII6}) means that the existence of a KCS is compatible 
\emph{only with type II models which are LRS} i.e. only within the invariant
subset $S_{1}^{+}(II)$. In order to determine the symmetry scale $\tilde{%
\alpha}$ we make use of the $\alpha \beta -$components of equation (\ref
{TypeII-1}). Then, from the spatial commutators (\ref{spatialcommutator}) we
get $X_{2,1}=X_{3,1}=0$ and the consistency of the remaining set of
equations is assured provided that: 
\begin{equation}
\tilde{\alpha}=\lambda \left( 4\Sigma _{22}+1\right) .  \label{TypeII7}
\end{equation}
We have proved that \emph{every type II evolving vacuum or }$\gamma -$\emph{%
law perfect fluid model that belongs to the invariant subset }$S_{1}^{+}(II)$%
\emph{\ can be invariantly characterized by the existence of a four
dimensional group of Kinematic Conformal Symmetries}.

Regarding the ``asymptotic behaviour'' of the KCS in type II models, it is
straightforward to show that in $S_{1}^{+}(II)|_{\Omega =0}$ the following
relations hold: 
\[
\tilde{\alpha}^{\prime }-\left( q+1\right) \tilde{\alpha}\propto \Sigma
_{22}^{2}-1 
\]
\[
\left( 3\tilde{\alpha}+\tilde{\delta}\right) ^{\prime }-\left( q+1\right)
\left( 3\tilde{\alpha}+\tilde{\delta}\right) \propto \Sigma _{22}^{2}-1. 
\]
Therefore, at the equilibrium points $\Sigma _{22}=\pm 1$ and $N_{1}=0$, the
KCS becomes a proper KSS as expected, since at the asymptotic regimes, all
models within $S_{1}^{+}(II)|_{\Omega =0}$ approach some vacuum Kasner model 
\cite{Wainwright-Hsu}.

On the other hand, it is well known that non-vacuum models $%
S_{1}^{+}(II)|_{\Omega >0}$ are all future asymptotic to the homothetic
Collins model $P_{1}^{+}(II)$ for which we have proved that does not admit a
proper KSS, and past asymptotic to a Kasner model $\mathcal{K}$ or to the
Friedmann-Lema\^{i}tre model $\mathcal{F}$. Consequently, every orbit in $%
S_{1}^{+}(II)|_{\Omega >0}$ joins two (first and second kind) self-similar
equilibrium points, hence we expect that the KCS will be reduced to a proper
HVF and a KSS, except to the case where the orbit lies in the
one-dimensional unstable manifold of $\mathcal{F}$. This model has past
attractor the point $\mathcal{F}$ and the KCS will reduce to the associated
KCS of the Robertson-Walker spacetime (equation (\ref{FRW5})).

Indeed, using equations (\ref{TypeII-2a})-(\ref{TypeII7}) we find: 
\[
\tilde{\alpha}-\tilde{\delta}=\frac{\lambda \left[ N_{1}^{2}+6\Sigma
_{22}\left( 2\Sigma _{22}-1\right) \right] }{3\Sigma _{22}}. 
\]
From the last equation we can show easily that $\left( \tilde{\alpha}-\tilde{%
\delta}\right) |_{P_{1}^{+}(II)}=0$ which implies that $\lim\limits_{\tau
\rightarrow +\infty }\left( \tilde{\alpha}-\tilde{\delta}\right) =0$.
Similarly we can show that, at the equilibrium point $\mathcal{K}$ the
symmetry scales $\alpha ,\delta $ become constants with $\alpha |_{\mathcal{K%
}}\neq \delta |_{\mathcal{K}}$ and at $\mathcal{F}$ we have $\alpha =\lambda
H $, $\lambda ^{\prime }=3\lambda $.

The exact form of the KCS in local coordinates is found to be: 
\begin{equation}
\mathbf{X}=\lambda (t)\partial _{t}+\left( 2cx+by\right) \partial
_{x}+cy\partial _{y}+\left( cz-1\right) \partial _{z}  \label{KCSTypeII-1}
\end{equation}
where $c$ is constant and: 
\begin{equation}
\lambda (t)=\frac{2cBC}{BC_{,t}-CB_{,t}}.  \label{KCSTypeII-2}
\end{equation}
The LRS type II metric is: 
\begin{equation}
ds^{2}=-A^{2}dt^{2}+B\left( dx+bzdy\right) ^{2}+C\left( dy^{2}+dz^{2}\right)
\label{KCSTypeII-3}
\end{equation}
for smooth functions $B(t),C(t)$ of their argument.

For example, in the LRS vacuum model (a special case of the general solution
found by Taub \cite{Taub1}) the metric is \cite{Wainwright-Ellis}: 
\begin{equation}
ds^{2}=-A^{2}dt^{2}+t^{2p_{1}}A^{-2}\left( dx+4p_{1}bzdy\right)
^{2}+t^{2p_{3}}A^{2}\left( dy^{2}+dz^{2}\right)  \label{KCSTypeII-4}
\end{equation}
where $A=\left( 1+b^{2}t^{4p_{1}}\right) ^{1/2}$ and $p_\alpha$ satisfy (\ref
{KasnerExponents}).

The temporal component of the KCS is given by: 
\begin{equation}
\lambda (t)=\frac{tA}{2tA_{,t}-\left( p_{1}-p_{3}\right) A}
\label{KCSTypeII-5}
\end{equation}
and reduces to a KSS for small and large values of the time coordinate (the
KSS in plane symmetric Bianchi models has also been found in \cite{Sintes1}).

As a final remark we note that by defining the new time coordinate $\tilde{t}%
-t_{0}=C(t)/B(t)$, the KCS becomes $\mathbf{X}=2c\tilde{t}\partial _{\tilde{t%
}}+\left( 2cx+by\right) \partial _{x}+cy\partial _{y}+\left( cz-1\right)
\partial _{z}$. Then at $\tilde{t}=0$ we have $\sigma _{ab}=0$ and equation (%
\ref{KCSwithmetric}) implies $c=0$ i.e. the KCS is reduced to the KVF of
type II models.

\subsection{Type VI$_{0}$ models}

Let us consider now the Bianchi type VI$_{0}$ invariant set which is defined
by $N_{1}=0$ and $N_{2}N_{3}<0$. Propagating equation (\ref{Habconstraint})
and using (\ref{expansion-shear}) we obtain:

\begin{eqnarray}
0 &=&\left( \lambda ^{\prime }-3\lambda \right) \Sigma _{\alpha \beta
}+\Lambda _{\alpha \beta }+\left[ 2\Sigma _{22}X^{\gamma }+\Sigma _{\hspace{%
0.2cm}\varepsilon }^{\gamma }X^{\varepsilon }\right] N_{2}\varepsilon
_{\gamma 2(\alpha }\delta _{\beta )}^{2}  \nonumber \\
&+&\left[ 2\Sigma _{33}X^{\gamma }+\Sigma _{\hspace{0.2cm}\varepsilon
}^{\gamma }X^{\varepsilon }\right] N_{3}\varepsilon _{\gamma 3(\alpha
}\delta _{\beta )}^{3}+\lambda S_{\alpha \beta }  \label{TypeVI0-1}
\end{eqnarray}
where $S_{\alpha \beta }$ and $\Lambda _{\alpha \beta }$ are given in (\ref
{spatial-curvature}) and (\ref{typeI5}) respectively.

The $\alpha \beta -$components of (\ref{TypeVI0-1}), after a short
calculation, give: 
\begin{equation}
\left( \Sigma _{22}-\Sigma _{33}\right) \left[ 2\partial _{\lbrack
3}X_{2]}+X_{1}\left( N_{2}+N_{3}\right) \right] =0  \label{TypeVI0-2}
\end{equation}
\begin{equation}
\left( 2\Sigma _{22}+\Sigma _{33}\right) \left( 2\partial _{\lbrack
2}X_{1]}+X_{3}N_{2}\right) =0=\left( 2\Sigma _{33}+\Sigma _{22}\right)
\left( -2\partial _{\lbrack 3}X_{1]}+X_{2}N_{3}\right)  \label{TypeVI0-3}
\end{equation}
which implies that $\Sigma _{22}=\Sigma _{33}$ since it can be shown that for $%
\Sigma _{22}\neq \Sigma _{33}$ the KCS is becoming a proper HVF.

In this case $N_{3}=-N_{2}$ and the system of equations (\ref{evol1})-(\ref
{Bianchi1}), (\ref{conformconstraint})-(\ref{Habconstraint}) and (\ref
{evolutalpha})-(\ref{evolutdelta}) is identically satisfied, provided that: 
\begin{equation}
\tilde{\alpha}=\lambda \left( 1-2\Sigma _{22}\right)  \label{TypeVI0-5}
\end{equation}
and the function $\lambda $ is given in terms of the shear and spatial
curvature variables ($\Sigma _{22}\neq 0$):

\begin{equation}
\lambda ^{\prime }=\frac{\lambda \left( 2N_{2}^{2}+9\Sigma _{22}\right) }{%
3\Sigma _{22}}.  \label{TypeVI0-6}
\end{equation}
Note that for $\Sigma _{22}=0$ the above system of equations implies $%
\lambda =0=\alpha =\delta$ and the KCS reduces to an isometry. Moreover, it
is not difficult to show that: 
\begin{equation}
\left( \tilde{\alpha}-\tilde{\delta}\right) |_{\Omega =0}=-\frac{2\lambda
\left( \Sigma _{22}+1\right) }{\Sigma _{22}}  \label{TypeVI0-7}
\end{equation}
\begin{equation}
\left( \tilde{\alpha}-\tilde{\delta}\right) |_{\Omega >0}=-\frac{2\lambda %
\left[ N_{2}^{2}+3\Sigma _{22}\left( \Sigma _{22}+1\right) \right] }{3\Sigma
_{22}}.  \label{TypeVI0-8}
\end{equation}
Due to equations (\ref{TypeVI0-7}) and (\ref{TypeVI0-8}), the analysis of
the asymptotic behaviour of the KCS is straightforward. The past and future
attractors of the vacuum invariant subset are the equilibrium points $\Sigma
_{22}=1,N_{2}=0$ (the arc $\mathcal{K}_{1}$) and $\Sigma _{22}=-1,N_{2}=0$
(the Taub point $\mathcal{T}_{1}$) respectively \cite{Wainwright-Ellis} and
the KCS reduces to a proper KSS (with $3\alpha +\delta =4\psi =0$) and a
proper HVF. Regarding the non-vacuum models it has been shown that the
future attractor is the Collins homothetic model $P_{1}^{+}(VI_{0})$ for
which $(\tilde{\alpha}-\tilde{\delta})|_{P_{1}^{+}(VI_{0})}=0$ and the KCS
reduces to a proper HVF. The past equilibrium state of the invariant set $%
S_{1}^{+}(VI_{0})$ is either the LRS Kasner point $\mathcal{Q}_{1}$ where
the KCS becomes a proper KSS or the Friedmann-Lema\^{i}tre model $\mathcal{F}
$ and the KCS is given in equation (\ref{FRW5}) with $\lambda ^{\prime
}=3\lambda $.

These results suggest that, \emph{every type VI}$_{0}$\emph{\ evolving
vacuum or }$\gamma -$\emph{law perfect fluid model lying in the invariant
subset }$S_{1}^{+}(VI_{0})$ \emph{always admits a four dimensional group of
KCS that reduces, at the asymptotic regimes, to a self-similarity group of
the first or second kind except from a set of measure zero for which the KCS
preserves its nature. }

For completeness we also give the form of the KCS for the general vacuum 
solution satisfying $N _{\hspace{0.2cm}\alpha }^{\alpha}=0$ (an arbitrary multiplication constant has
been omitted) \cite{Ellis-MacCallum, Exact-Solutions-Book}: 
\begin{equation}
ds^{2}=t^{-1/2}e^{t^{2}}\left( -dt^{2}+dx^{2}\right) +t\left(
e^{2x}dy^{2}+e^{-2x}dz^{2}\right)
\end{equation}
\begin{equation}
\mathbf{X}=\frac{4t\left( c_{1}+c_{2}\right) }{4t^{2}-3}\partial
_{t}+c_{1}\partial _{x}+c_{2}y\partial _{y}+\left( 2c_{1}+c_{2}\right)
z\partial _{z}  \label{TypeVI0-10}
\end{equation}
where the symmetry scales are:\ 
\begin{equation}
\alpha =\frac{\left( c_{1}+c_{2}\right) \left( 4t^{2}-1\right) }{\left(
4t^{2}-3\right) },\qquad \delta =\frac{\left( c_{1}+c_{2}\right) \left(
4t^{2}-9\right) \left( 4t^{2}+1\right) }{\left( 4t^{2}-3\right) ^{2}}.
\label{TypeVI0-11}
\end{equation}
We can verify that, at early and late times, we have $3\alpha +\delta
\approx 0$ and $\alpha -\delta \approx 0$, so the expected limiting cases of
the KCS is a self-similarity of the second or first kind\footnote{%
We recall here that the time coordinate $t$ does not represent the proper
clock time as $t\rightarrow 0^{+}$ or $t\rightarrow +\infty $. However, the
freedom of choosing the time gauge, ensures a similar asymptotic behaviour
for the symmetry scales.}. Also to be noted is the apparent ``singular''
behaviour (due to the specific time gauge) of the KCS as $t\rightarrow \sqrt{%
3}/2$ ($\Rightarrow \sigma _{ab}\rightarrow 0$). However, similarly to the
previous case, we can show that at this value, the temporal component of $%
\mathbf{X}$ vanishes and the symmetry equations imply that $c_{1}+c_{2}=0$
i.e. the KCS $\mathbf{X}\rightarrow \partial _{x}-y\partial _{y}+z\partial
_{z}$ reduces to the KVF of Bianchi type VI$_{0}$ models.

\section{Discussion}

As we have mentioned, it is customary to study a specific symmetry
assumption from a geometrical point of view without taking into
consideration the kinematical and dynamical structure of the corresponding
model. As a consequence, the effects on the dynamics of the symmetry
constraints are hidden and usually produce unphysical results. Up to date,
self-similarity of the first kind appears to be the only symmetry condition
with a transparent physical nature since it represents \emph{a
geometrization of the asymptotic (equilibrium) state of general models}.
Therefore, in order to complete the dynamical picture, it was of interest to
seek and find a symmetry that could be used effectively as a consistent tool
for the invariant description of the intermediate behaviour of general
models.

In the present article we have proposed a new technique of studying
geometric symmetries by fully exploiting the 1+3 orthonormal frame scheme
and the introduction of expansion-normalized variables. The analysis of the
complete set of consistency equations (\ref{evol1})-(\ref{Bianchi1}), (\ref
{conformconstraint})-(\ref{Habconstraint}) and (\ref{evolutalpha})-(\ref
{evolutdelta}) has revealed a \emph{novel feature} of a large class of SH
models that is summarized in the following:\newline
\newline
\textbf{Proposition 1} \emph{Evolving SH models of Bianchi type I, LRS models
of Bianchi type II and type VI$_{0}$ models within the associated invariant subset} $N_{%
\hspace{0.2cm}\alpha }^{\alpha }=0$\emph{, are geometrically described by
the existence of a four dimensional group of KCS that is reduced to a
self-similarity group of the first or second kind at the asymptotic regimes,
except from a set of measure zero for which the properness of the KCS is
preserved.}\newline

However, as we have seen, even in the exceptional cases the corresponding
equilibrium state is the Friedmann-Lema\^{i}tre model $\mathcal{F}$ which we
have proved that always admits a proper KCS, supplementing the geometric
properties of the standard cosmological model.

At first sight, the existence of a proper KCS in SH models appears somewhat
surprising, at least from a dynamical point of view. This is mainly because
the interaction mechanism between the geometric ``assumption" of a KCS and
the dynamical behaviour of SH models is not, conceptually, apparent. However
a closer look on the structural properties of SH models indicates a possible
qualitative interpretation of this interaction. In particular, the full set
of non-linear EFE can be seen as the perturbed version of the associated
linearization of equations (\ref{evol1})-(\ref{Bianchi1}), at the vicinity
of a hyperbolic equilibrium point. Accordingly we may interpret the
generator of a KCS as representing the perturbation (to some order) of the
corresponding generator of the self-similarity transformation group of the
first or second kind. Eventually, this observation will enable us to
geometrize the majority of the concepts and techniques that are used in the
theory of dynamical systems with a view to optimize and efficiently
elaborate the results from the qualitative study of general cosmological
models.

Although the existence of a KCS signifies the physical ground for a first
promising attempt towards a ``geometrisation'' of the evolutionary behaviour
of SH models, we do not allegate that the KCS (uniquely) characterizes the intermediate epoch of
the totality of SH models. In fact, because a KCS possess two arbitrary spatially homogeneous functions 
(the symmetry scales), we expect that \emph{only} SH models with \emph{two} essential degrees of freedom will exhibit a proper KCS. 
This conjecture is confirmed by the Proposition 1 in which all the exact solutions found to admit a proper KCS belong to this class. 
The main reason, that provides an interpretation for the possible connection between exact solutions and the existence of KCS, appears to involve the so-called 
``hidden" symmetries of the SH cosmologies. Indeed, using the Hamilton-Jacobi reformulation of the EFE in which all the dynamical picture is encoded in one geometric object, 
it has been shown that all known exact solutions with two degrees of freedom are associated with a specific kind of ``hidden" symmetry namely the existence 
of a Killing tensor symmetry of the Jacobi metric that generalizes the corresponding cyclic variables and the Hamilton-Jacobi separability 
\cite{Uggla-Rosquist-Jantzen1, Uggla-Rosquist-Jantzen2}. Therefore it will be of interest to extent the analysis to the rest of the Bianchi models 
in order to see if the existence of a KCS is a general feature of the two-dimensional SH invariant subset 
i.e. if it is related with the above type of ``hidden" symmetry of the Bianchi cosmological models \cite{Apostol-Carot1}.  

We should remark that, although we have mainly focused our study to Bianchi types with clear and simple past and future equilibrium states, 
one could apply the approach presented in this paper to models with more complicated dynamical structure e.g. models with oscillating or diverging 
asymptotic behaviour near to the past or future attractor. The non-vacuum Bianchi VII$_0$ invariant subset provides an interesting example since it is well known that the 
associated kinematical variables are unbounded and do not approach any equilibrium point into the future i.e. 
non-vacuum Bianchi type VII$_0$ models are not asymptotically self-similar \cite{Wainwright-Ellis, Hervik-Hoogen-Lim-Coley}. As a consequence one should expect that a KCS does not exist 
in those models due to the non-existence of a self-similar model as future attractor. However, a preliminary analysis has shown that 
a KCS does exist in LRS type VII$_0$ models which is never reduced to a HVF or a KSS \cite{Apostol-Carot1}, suggesting that the concept of the KCS represents 
not simply a perturbed version of the self-similarity group but a generic property (in the spirit of \cite{Uggla-Rosquist-Jantzen1, Uggla-Rosquist-Jantzen2}) 
of the two-dimensional invariant subset of the SH cosmological models. 

Clearly, the case of higher dimensional invariant subsets requires further investigation. Assuming the existence of several proper KCS, will not solve the problem for models 
with three or more essential degrees of freedom, since the condition (\ref{definitionKCS}) implies that the symmetry scales are always spatially homogeneous restricting the 
dimension of the Lie algebra of KCS in SH models to four. Therefore the question of determining the symmetry which invariantly describes the whole set of SH models is 
still open. Nevertheless, the implications of the above results, enforce the important role which may play a specific (still unknown) general symmetry as an 
effective geometric implement for the invariant description of general cosmological models and not only as a simplification rule towards the determination of exact solutions 
with ambiguous (or even without any) physical meaning. A closely related issue is how the constraints, coming from the presence of the general symmetry, could reveal a path of
constructing the general (whenever is possible) solution of the corresponding cosmological model. We expect that, the approach of studying generic geometric symmetries 
in spacetime developed in this article, can be applied to more general geometric setups leading to a more efficient qualitative and analytical study of general vacuum 
and perfect fluid models. 

\vskip 0.5cm

\centerline{\bf\large Acknowledgments} \vskip .3cm The author would like to thank the anonymous referees for suggesting 
changes that have improved the presentation of the manuscript. This work is supported
through the research programme ``Pythagoras'' of the Greek Ministry of National 
Education, contract No 70-03-7310. Also the author gratefully acknowledges the award of a postdoctoral fellowship from the 
Spanish ``Ministerio de Educaci\'on y Ciencia", grant No SB2004-0110.

\end{document}